\documentclass[aps,prd,showpacs,superscriptaddress,nofootinbib,floatfix,10pt]{revtex4-2} 

\usepackage[utf8]{inputenc}
\usepackage{color}
\usepackage{amsfonts,amsmath,amssymb}
\usepackage{longtable,booktabs}
\usepackage{graphicx}
\usepackage{bm}
\usepackage{xcolor}

\usepackage[colorlinks=true,linkcolor=blue,citecolor=blue]{hyperref}
\usepackage{epstopdf}
\usepackage{mathtools}
\usepackage{subfigure}
\usepackage{multirow}
\usepackage{diagbox}
\usepackage{afterpage}
\usepackage{placeins}
\usepackage{float}
\usepackage{mathrsfs}
\usepackage{orcidlink}
\usepackage{tabularray}
\usepackage{geometry}
\allowdisplaybreaks[1]

\definecolor{zsgreen}{RGB}{6,119,0}
\definecolor{zsorg}{RGB}{253,112,28}
\definecolor{zsorgd}{RGB}{207,85,12}

\newcolumntype{M}[1]{>{\centering\arraybackslash}m{#1}}
\newcolumntype{N}{@{}m{0pt}@{}}
\allowdisplaybreaks[1]
\graphicspath{{figs/}}

\newcommand{\itp}{\affiliation{Institute of Theoretical Physics, Chinese Academy of Sciences, Beijing 100190, China}}

\newcommand{\qfnu}{\affiliation{College of Physics and Engineering, Qufu Normal University, Qufu 273165, China}}

\newcommand{\chep}{\affiliation{Center for High Energy Physics, Peking University, Beijing 100871, China}}

\begin{document}

\title{Role of electromagnetic corrections in the \texorpdfstring{$\pi\pi$}{pipi} distributions of  \texorpdfstring{$\psi^\prime \to J/\psi \pi \pi$}{psiprimetoJpsipipi}}

\author{Zhao-Sai Jia\orcidlink{0000-0002-7133-189X}}\qfnu \itp

\author{Gang Li\orcidlink{0000-0002-5227-8296}}\email{gli@qfnu.edu.cn} \qfnu

\author{Zhen-Hua Zhang
\orcidlink{0000-0001-6072-5378}}\email{zhhzhang@pku.edu.cn}\chep

\begin{abstract}
The cusp structure at the $\pi^+\pi^-$ threshold in the $\pi^0\pi^0$ invariant mass spectrum serves as a sensitive probe for extracting the $S$-wave $\pi\pi$ scattering lengths in processes where an $S$-wave $\pi^0\pi^0$ pair is produced in the final states. Within the framework of nonrelativistic effective field theory with coupled channels $\pi^0\pi^0$ and $\pi^+\pi^-$, we revisit the near-threshold structures in the $\pi^0\pi^0$ spectrum of $\psi^\prime \to J/\psi \pi\pi$. Our analysis incorporates the $\pi \pi$ final-state rescattering, including both strong and Coulomb interactions. It turns out that the cusp near the $\pi^+\pi^-$ threshold becomes more prominent when Coulomb interactions are included. The electromagnetic correctionsare found to alter the magnitude of the threshold cusp by about 2\%-3\%, underscoring the necessity of including these effects in precision determinations of the $\pi\pi$ scattering lengths. The coupled-channel amplitude constructed in this work provides a ready-to-use theoretical framework for experimental analyses of fine structures near $\pi\pi$ thresholds.
\end{abstract}

\maketitle

\section{Introduction}

As the lightest quantum chromodynamics (QCD) bound states and the pseudo-Nambu-Goldstone bosons of spontaneous chiral symmetry breaking, pions play an essential role in understanding low-energy strong interaction dynamics. At low energies, the $\pi\pi$ scattering lengths are key to quantifying the strength of the $\pi\pi$ $S$-wave interaction, which can shed light on essential information about chiral symmetry breaking. The $\pi\pi$ scattering lengths have been extensively studied both theoretically and experimentally. Theoretically, the isospin $I=0,\,2$ scattering lengths and their difference were predicted with percent-level accuracy in units of the inverse pion mass $M_{\pi}^{-1}$~\cite{Colangelo:2000jc, Colangelo:2001df}, $a_0 M_{\pi^+}=0.220 \pm 0.005$, $a_2 M_{\pi^+}=-0.0444 \pm 0.0010$, and $(a_0-a_2) M_{\pi^+}=0.265 \pm 0.004$, by combining chiral perturbation theory (ChPT) at the two-loop level~\cite{Weinberg:1978kz, Gasser:1983yg} and the Roy equations~\cite{Roy:1971tc}, consistent with dispersion relation approaches without input from ChPT~\cite{Garcia-Martin:2011iqs}.

Experimentally, the $\pi\pi$ scattering lengths is extracted mainly through three approaches. First, using Watson's theorem and numerical solutions of the Roy equations~\cite{Ananthanarayan:2000ht,Descotes-Genon:2001oav}, it is possible to determine the $\pi\pi$ scattering lengths from the angular distributions of $K_{e4}$ decay. Experiments along these lines have been carried out in, e.g., Refs.~\cite{Rosselet:1976pu,Pislak:2003sv,NA482:2007xvj,NA482:2010dug}. 
Second, the lifetime of pionium (the $\pi^+\pi^-$ hadronic atom) that decays into two neutral pions is related to $|a_0-a_2|$~\cite{Palfrey:1961kt}. The experimental results reported by the DIRAC Collaboration are well consistent with the ChPT prediction~\cite{DIRAC:2005hsg}. 
Third, the cusp effect in decay processes with at least two neutral pions in the final states, such as $K^{\pm} \to \pi^0\pi^0\pi^{\pm}$~\cite{Budini:1961bac,Cabibbo:2004gq,Cabibbo:2005ez,Colangelo:2006va,NA482:2005wht,Batley:2009ubw,Gasser:2011ju,Gevorkyan:2013tsa}, $K_L \to 3\pi$~\cite{Bissegger:2007yq}, $\eta \to 3\pi$~\cite{Ditsche:2008cq}, and $\eta^\prime \to \eta \pi\pi$~\cite{Kubis:2009sb}, has emerged as a precise tool for extracting $\pi\pi$ scattering lengths, which results from the charge exchange rescattering $\pi^+\pi^- \to \pi^0\pi^0$~\cite{Meissner:1997fa}. The cusp effect is prominently manifested in the $\pi^0\pi^0$ invariant mass spectrum of the decay $K^{\pm} \to \pi^0\pi^0\pi^{\pm}$. 
The branching ratio difference $\mathcal{B}[K^{\pm} \to \pi^{\pm} \pi^{\mp} \pi^{\pm}]/\mathcal{B}[K^{\pm} \to \pi^0 \pi^0 \pi^{\pm}]=(5.583\pm0.024)\%/(1.760\pm0.023)\%$~\cite{ParticleDataGroup:2024cfk} requires significant charge exchange rescattering contributions. The complexity of final state interactions in the aforementioned decay processes, together with the reconstruction of the $\eta$, which decays into $2\gamma$ or $3\pi^0$, in $\eta$ decay processes, constitute challenges for studying the cusp effect in $\eta'$ decays.
Combining both the $K_{e4}$~\cite{NA482:2010dug} and $K\to 3\pi$~\cite{Batley:2009ubw} measurements, the NA48/2 experiment reported 
$a_0^0M_{\pi^+}=0.2210 \pm 0.0047_{\mathrm{stat}} \pm 0.0040_{\mathrm{syst}}$, $a_0^2M_{\pi^+}=-0.0429 \pm 0.0044_{\mathrm{stat}} \pm 0.0028_{\mathrm{syst}}$, and $(a_0^0-a_0^2)M_{\pi^+}=0.2639 \pm 0.0020_{\mathrm{stat}} \pm 0.0015_{\mathrm{syst}}$~\cite{NA482:2010dug}.

In this regard, heavy quarkonium dipion transitions have also been considered for extracting the $\pi \pi$ $S$-wave phase shifts~\cite{Guo:2005uxv,Zhang:2008tm,Chen:2009zzr} and scattering lengths~\cite{Liu:2012dv}. In these processes, the interaction between the final-state quarkonium and pions is highly suppressed by the Okubo-Zweig-Iizuka (OZI) rule, and thus their contributions can be neglected, as indicated by the lattice result for the $J/\psi$ scattering length $a_{J/\psi\pi}=(-0.01\pm 0.01)$~fm~\cite{Liu:2008rza}. 
The dipion transitions also exhibit sizable branching fractions of heavy quarkonium decays. For example, the branching fraction of $\psi^\prime \to J/\psi \pi^0\pi^0$ is $(18.2 \pm 0.5)\%$~\cite{ParticleDataGroup:2024cfk}. One also has $\mathcal{B}(\psi^\prime \to J/\psi \pi^+\pi^-)/\mathcal{B}(\psi^\prime \to J/\psi \pi^0\pi^0) \approx 2$~\cite{ParticleDataGroup:2024cfk}, due to the approximate isospin symmetry. 
The possibility of extracting the $\pi\pi$ scattering lengths using the cusp effect in the heavy quarkonium dipion transition $\Upsilon(3S) \to \Upsilon(2S) \pi^0\pi^0$ was studied in Ref.~\cite{Liu:2012dv} within the nonrelativistic effective field theory (NREFT) framework~\cite{Gasser:2011ju,Kubis:2009sb}. Using Monte Carlo (MC) simulation, the statistical precision of $a_0-a_2$ was estimated to reach 2\% and 1.5\% with $2\times 10^7$ and $4\times 10^7$ events of $\Upsilon(3S) \to \Upsilon(2S) \pi^0\pi^0$, respectively. However, the Coulomb interaction was not included in the framework of Ref.~\cite{Liu:2012dv}. 
The Coulomb interactions between $\pi^+$ and $\pi^-$ have been incorporated in the $K^{\pm} \to \pi^0\pi^0\pi^{\pm}$ process by the NA48/2 Collaboration~\cite{Gevorkian:2006rad,Batley:2009ubw}, where the $\pi \pi$ scattering lengths were defined to be proportional to the $\pi \pi$ scattering amplitude at the threshold in the isospin limit. 
Radiative corrections can reduce $a_0-a_2$ and $a_0$ by about 9\% and 6\%, respectively. Therefore, the Coulomb interactions could also give non-negligible contributions in heavy quarkonium dipion transitions.

Huge data samples have already been accumulated at the BESIII experiment in the $\psi^\prime \to J/\psi \pi^0\pi^0$ channel. In particular, a sample of $2.7\times 10^9$ $\psi^\prime$ events has been collected at the BESIII experiment~\cite{BESIII:2024lks}, and $6.4\times 10^{11}$ $\psi'$ events can be collected with one year of running at the Super Tau-Charm Facility (STCF)~\cite{Achasov:2023gey} under consideration.
It is therefore of great interest to have an accurate framework for analyzing the near-threshold $\pi\pi$ invariant mass distributions of $\psi^\prime \to J/\psi \pi\pi$.
This is the goal of this work, with an emphasis on investigating the cusp effect in the $\pi^0\pi^0$ invariant mass distribution of $\psi^\prime \to J/\psi \pi^0\pi^0$ within the NREFT framework. The $\pi\pi$ final state interactions (FSIs) are included along with the Coulomb interactions between the $\pi^+\pi^-$ pair, while the $J/\psi \pi$ FSI will be neglected. 
Using inputs from the $\pi^0\pi^0$ scattering lengths, the energy level and lifetime of the ground-state pionium, and the BESII and ATLAS data~\cite{BES:2006eer,ATLAS:2016kwu} of the $\pi^+\pi^-$ spectrum for $\psi^\prime \to J/\psi \pi^0\pi^0$, we predict the near-threshold line shapes of the $\pi^0\pi^0$ and $\pi^+\pi^-$ invariant mass distributions. We also perform a MC simulation to provide a quantitative estimation of the effects of electromagnetic corrections on the experimental extraction of the $\pi\pi$ scattering lengths. 

The structure of the paper is as follows: In Sec.~\ref{Sec:NREFT}, we derive expressions for the $\psi^\prime \to J/\psi \pi\pi$ partial decay width within the coupled-channel NREFT. In Sec.~\ref{Sec:MC}, we generate synthetic data using MC simulation to estimate the precision required for the extraction of scattering lengths when electromagnetic corrections should be considered. 
In Sec.~\ref{Sec:Results}, the near-threshold line shapes of the $\pi\pi$ invariant mass distribution are predicted for the $\psi^\prime \to J/\psi \pi\pi$ process. A brief summary is given in Sec.~\ref{Sec:Summary}.

\section{Non-relativistic effective field theory}\label{Sec:NREFT}
\begin{figure*}[tb]
    \centering
    \includegraphics[width=0.9\linewidth]{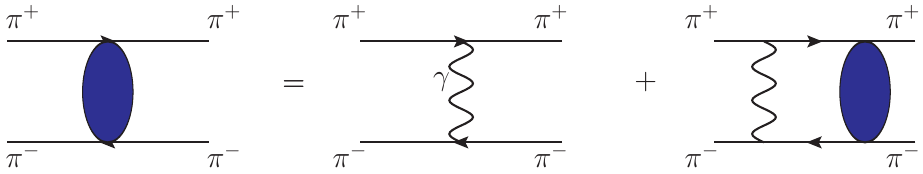}
    \caption{Resummation in the Coulomb $T$ matrix for the $\pi^+\pi^-$ scattering. The wavy line represents a photon, and the blue blob represents resummation of the Coulomb photon exchanges.}
    \label{fig:Coulomb_resum}
\end{figure*}
In this section, we present the amplitudes and differential decay widths of $\psi^\prime \to J/\psi \pi\pi$ near the $\pi^+\pi^-$ threshold. In the following, the $\pi^0\pi^0$ and $\pi^+\pi^-$ channels are designated as channels 1 and 2, respectively.

\subsection{\texorpdfstring{$\pi\pi$}{pipi} \texorpdfstring{$S$}{S}-wave scattering amplitude}

The energy dependence of the differential decay widths near the threshold is predominantly determined by the $\pi\pi$ final-state interaction amplitudes, which receive contributions from both Coulomb and strong interactions. In this work, we focus on $S$-wave $\pi\pi$ scattering, since the near-threshold $D$-wave amplitude is suppressed by a factor of $v_\pi^2$ relative to the $S$-wave amplitude, where $v_\pi$ is the pion velocity in the center-of-mass (c.m.) frame. The $S$-wave two-channel scattering $T$ matrix can be expressed in the two-potential formalism as~\cite{Braaten:2017kci,Shi:2021hzm,Jia:2024ybo}
\begin{align}
    &\bm{T}(E) = \begin{pmatrix}
 0 & 0 \\
 0 & T_C(E)
\end{pmatrix} 
+ \begin{pmatrix}
    1 & 0 \\
    0 & W_C(E) \end{pmatrix} 
    \bm{T}_{\rm SC}(E) 
    \begin{pmatrix}
    1 & 0 \\
    0 & W_C(E) \end{pmatrix},
    \label{Eq_Tmatrix}
\end{align}
where $E=\sqrt{s}$ is the total energy in the $\pi\pi$ c.m. frame,
\begin{align}
    T_C(E) = \frac{i \pi}{ \mu_c k_c} \left(\frac{\Gamma(1-ix)}{\Gamma(1+ix)} - 1\right),
    \label{Eq_TC}
\end{align}
is the $\pi^+\pi^-$ Coulomb scattering amplitude including infinite Coulomb photon exchanges as shown in Fig.~\ref{fig:Coulomb_resum}, with $\mu_c=M_{\pi^+}/2$ the $\pi^+\pi^-$ reduced mass, $k_c=\sqrt{2\mu_c(E-2M_{\pi^+})}$ the c.m. momentum of $\pi^+$, $x=\alpha\mu_c/k_c$, and $\alpha=1/137$ the fine-structure constant; 
and
\begin{align}
    W_C(E)=\left(\frac{2 \pi x}{1-e^{-2 \pi x}} \frac{\Gamma(1-i x)}{\Gamma(1+i x)}\right)^{{1}/{2}},
    \label{Eq_WC}
\end{align} 
accounts for the Coulomb photon exchanges between the $\pi^+\pi^-$ pair in the initial or final states. $\bm{T}_{\rm SC}$ is the strong-Coulomb scattering amplitude. For the $S$-wave $\pi\pi$ near-threshold scattering, at leading order (LO) in NREFT, there are only constant contact terms for the strong interaction-induced pion-pion scattering potential, and $\bm{T}_{\rm SC}$ satisfies the Lippmann-Schwinger equation (LSE),
\begin{align}
    \bm{T}_{\rm SC}(E) 
    &=\left[\bm{I}-\bm{V}_S(\Lambda)\bm{G}_C(E, \Lambda)\right]^{-1}\bm{V}_S(\Lambda),
 \label{Eq_TSCmatrix}
\end{align}
where $\bm{V}_S$ is the strong potential, $\bm{G}_C(E, \Lambda)=\mathrm{diag}(G_{C11},G_{C22})$ is the Green's function regularized by a sharp cutoff $\Lambda$, with nonvanishing diagonal matrix elements~\cite{Kong:1999sf,Konig:2015aka,Braaten:2017kci}
\begin{align}
    G_{C 11}(E, \Lambda) &= -\frac{\mu_n \Lambda}{\pi^2}-i \frac{\mu_n}{2 \pi} k_n(E), \label{eq:GC11}\\
    G_{C 22}(E, \Lambda) &= -\frac{\mu_c \Lambda}{\pi^2}-\frac{\alpha \mu_c^2}{\pi}\left(\ln \frac{\Lambda}{\alpha \mu_c}-\gamma_E\right)-\frac{\mu_c}{2 \pi} \kappa_c(E). \label{eq:GC22}
\end{align}
Here, $\mu_n=M_{\pi^0}/2$ is the $\pi^0\pi^0$ reduced mass, $\gamma_E$ is the Euler constant, $k_n=\sqrt{2\mu_n(E-2M_{\pi^0})}$ is the c.m. momentum of $\pi^0$, and $\kappa_c = 2\alpha\mu_c\left[\ln(ix)+1/(2ix)-\psi(-ix)\right]$, with $\psi(x)$ being the digamma function.

Since the strong-Coulomb scattering amplitude $\bm{T}_{\rm SC}(E)$ is independent of the cutoff $\Lambda$, the strong potential $\bm{V}_S$ in the LSE must be $\Lambda$ dependent to absorb the $\Lambda$ dependence in $\bm{G}_C(E,\Lambda)$. After renormalization, $\bm{T}_{\rm SC}(E)$ can be expressed in terms of cutoff-free parameters ~\cite{Kong:1999sf,Jia:2024ybo}
\begin{align}
    \bm{T}_{\rm SC}^{-1}(E) 
   &=(\bm{V}_S^R)^{-1} - \bm{G}_C^R(E)\nonumber\\
   &=(\bm{T}_{SC}^{\text{thr}})^{-1} -\tilde{\bm{G}}_{C}^R(E),
    \label{Eq_TSCsl}
\end{align}
where
\begin{align}
    \bm{G}_C^R(E) = \begin{pmatrix}
        \frac{-i\mu_n}{2\pi} k_n(E) & 0 \\
        0 & \frac{-\mu_c}{2\pi} \kappa_c(E)
    \end{pmatrix}
\end{align}
is the renormalized Green's function, $\tilde{\bm{G}}_{C}^R=\bm{G}_{C}^{R}(E) - \bm{G}_C^R(E=2M_{\pi^0})$, and \begin{align}
    \bm{T}_{SC}^{\rm{thr}}=-2\pi \bm{\mu}^{-\frac{1}{2}} \bm{a}_{NN, \, \rm{eff}} \bm{\mu}^{-\frac{1}{2}}
\end{align}
represents the $\bm{T}_{SC}$ at the $\pi^0\pi^0$ threshold expressed in terms of the $\pi\pi$ scattering lengths~\cite{Sakai:2020psu} with $\bm{\mu} =$ diag($\mu_n, \mu_c$). Here 
\begin{align}
    \bm{a}_{NN,\rm{eff}} = \begin{pmatrix}
        a_{11, \, \text{eff}} & a_{12, \, \text{eff}} \\
        a_{12, \, \text{eff}} & a_{22, \, \text{eff}}
    \end{pmatrix}
\end{align}
denotes the scattering length matrix, and its matrix elements can be written in terms of the isospin $I=0$ and $I=2$ $S$-wave $\pi\pi$ scattering lengths $a_0$ and $a_2$ as

\begin{align}
    a_{11, \, \text{eff}} &= \frac{2}{3M_{\pi^0}} (a_0+2a_2) M_{\pi^+} (1-\eta),\label{Eq_a11eff} \\
    a_{12, \, \text{eff}} &= \frac{2}{3M_{\pi^0}} (a_0-a_2) M_{\pi^+} (1+\frac{\eta}{3}),\nonumber \\
    a_{22, \, \text{eff}} &= \frac{1}{3M_{\pi^+}} (2a_0+a_2) M_{\pi^+} (1+\eta),
    \label{Eq_a12eff}
\end{align}
where $\eta=(M_{\pi^+}^2-M_{\pi^0}^2)/M_{\pi^+}^2$ accounts for the isospin breaking effect at LO in ChPT~\cite{Liu:2012dv,Gasser:2011ju}, and $(a_0-a_2) M_{\pi^+}=0.2640 \pm 0.004$, $a_2 M_{\pi^+}=-0.0444 \pm 0.0010$~\cite{Gasser:2009zz,Colangelo:2000jc,Colangelo:2001df}.

\subsection{\texorpdfstring{$\pi\pi$}{pipi} invariant mass distribution in \texorpdfstring{$\psi^{\prime}\to J/\psi \pi\pi$}{psi'->J/psi pi pi}}
\begin{figure*}[tb]
    \centering
    \includegraphics[width=0.68\linewidth]{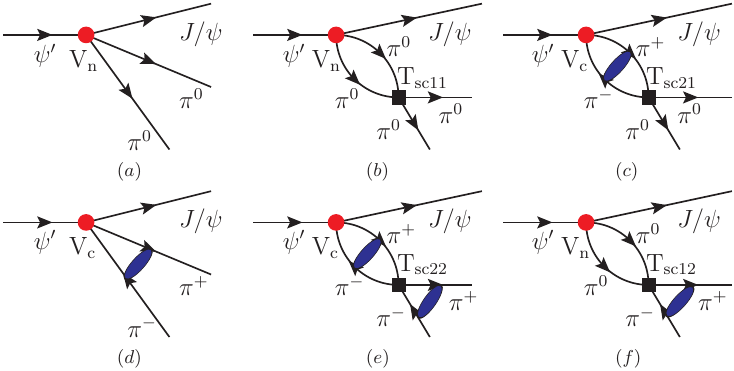}
      \caption{Diagrams for $\psi^\prime \to J/\psi \pi^0\pi^0$ (a-c) and $\psi^\prime \to J/\psi \pi^+\pi^-$ (d-f). The red filled circles represent the pointlike production sources $\mathrm{V_n}$ and $\mathrm{V_c}$ for $J/\psi \pi^0\pi^0$ and $J/\psi \pi^+\pi^-$ production, respectively. The black filled squares represent the $\pi\pi$ interactions described by the solution of the LSE in Eq.~\eqref{Eq_TSCsl}. The blue blob represents infinite Coulomb-photon exchanges between $\pi^+$ and $\pi^-$ as shown in Fig.~\ref{fig:Coulomb_resum}. The Coulomb-photon resummation between the final-state $\pi^+$ and $\pi^-$ is accounted for by the $W_C$ factor in the amplitude for $\psi^\prime \to J/\psi \pi^+\pi^-$ given in Eq.~\eqref{Eq_Mcc}. The amplitude for $\psi^\prime \to J/\psi \pi^0\pi^0$ is given in Eq.~\eqref{Eq_Mnn}.}
    \label{Figs_Feynmandiagrams}
\end{figure*}

The diagrams illustrating the $\psi^\prime \to J/\psi \pi\pi$ processes are shown in Fig.~\ref{Figs_Feynmandiagrams}. In the near-threshold region, the short-distance decay vertices for $\psi^\prime \to J/\psi \pi\pi$ can be approximated as constants, and the decay amplitudes can be written as
\begin{align}
    \mathcal{M}[\psi^\prime \to J/\psi \pi^0\pi^0]  &= \mathrm{V_n^\Lambda} + \mathrm{V_n^\Lambda} \times \frac{1}{2}G_{C11}^\Lambda \times T_{\rm SC11} + \mathrm{V_c^\Lambda} \times G_{C22}^\Lambda \times T_{\rm SC21},\label{Eq_Mnn}\\
    \mathcal{M}[\psi^\prime \to J/\psi \pi^+\pi^-] &= \sqrt{2}\left(\mathrm{V_c^\Lambda} + \mathrm{V_c^\Lambda} \times G_{C22}^\Lambda \times T_{\rm SC22}
    + \mathrm{V_n^\Lambda} \times \frac{1}{2}G_{C11}^\Lambda \times T_{\rm SC12}\right) \times W_C,\label{Eq_Mcc}
\end{align}
where $\mathrm{V_n^\Lambda}$ and $\mathrm{V_c^\Lambda}$ are the cutoff-dependent short-distance vertices for $\psi^\prime \to J/\psi \pi^0\pi^0$ and $\psi^\prime \to J/\psi \pi^+\pi^-$, with the final-state $\pi\pi$ pair in the $S$-wave. The $1/2$ factors in the equations account for the effects of identical neutral pions in the intermediate loops, and the effect of identical pions in the $J/\psi\pi^0\pi^0$ final state is accounted for by the $\sqrt{2}$ factor in Eq.~\eqref{Eq_Mcc} relative to Eq.~\eqref{Eq_Mnn}. The $\Lambda$ dependence in the intermediate Green's functions in Eqs.~\eqref{Eq_Mnn} and \eqref{Eq_Mcc} can be absorbed by the short-distance vertices 
$\mathrm{V_n^\Lambda}$ and  $\mathrm{V_c^\Lambda}$ through multiplicative renormalization~\cite{Braaten:2005jj,Sakai:2020psu}. The cutoff-independent amplitudes obtained by keeping only the LO terms in the Green's functions are
\begin{align}
    \mathcal{M}[\psi^\prime \to J/\psi \pi^0\pi^0]  &= \mathrm{V_n} \times \left(\frac{1}{2} T_{\rm SC11} + T_{\rm SC21}\right),\label{Eq_MnnLO}\\
    \mathcal{M}[\psi^\prime \to J/\psi \pi^+\pi^-] &= \mathrm{\sqrt{2}V_n} \times \left(T_{\rm SC22} + \frac{1}{2} T_{\rm SC12}\right) \times W_C,\label{Eq_MccLO}
\end{align}
where $\mathrm{V_n}$ is the renormalized $\Lambda$-independent short-distance vertex. We have taken the isospin limit in the renormalization so that $\mathrm{V_c}\simeq\mathrm{V_n}$ and $\mu_n \simeq \mu_c$ in the linearly divergent terms of the Green's functions. 

In Refs.~\cite{Dong:2021lkh, Wu:2024xwy}, the short-distance vertex for $\psi^\prime \to J/\psi \pi\pi$ was derived using ChPT~\cite{Mannel:1995jt}, and the parameters in the vertex were fixed by fitting to the BESII and ATLAS data~\cite{BES:2006eer,ATLAS:2016kwu} for the $\pi^+\pi^-$ spectrum and the helicity angular distribution, with the $\pi\pi$ FSIs included through the Omn\`{e}s function.
By matching the $\psi^\prime \to J/\psi \pi\pi$ vertex from the LO NREFT to that from ChPT in Refs.~\cite{Wu:2024xwy,Dong:2021lkh} at the $\pi\pi$ threshold, one obtains
\begin{align}
    \mathrm{V_n} =& -\frac{\sqrt{6m_{\psi^\prime} m_{J/\psi}}M_{\pi}  \alpha^{21} } {\beta^0}\left\{(9\kappa-10) M_{\pi}^2+ 3\kappa\left[ \frac{\lambda(m_{\psi^\prime}^2,4M_{\pi}^2,m_{J/\psi}^2)}{4 m_{\psi^\prime}^2}\right]\right\},
\end{align}
where $m_{\psi^\prime}$ and $m_{J/\psi}$ are the masses of $\psi^\prime$ and $J/\psi$, respectively, $\lambda(a,b,c)=a^2+b^2+c^2-2ab-2ac-2bc$ is the K\"all\'en function,
$\beta^0=(33-2N_f)/3$ is the first coefficient of the QCD beta function~\cite{Brambilla:2015rqa,Voloshin:2007dx} with $N_f$ being the number of light-quark flavors, $|\alpha^{21}|=(1.18 \pm 0.01 \pm 0.05)~\rm{GeV^{-3}}$, and $\kappa=0.26 \pm 0.01 \pm 0.01$~\cite{Wu:2024xwy}.

The $\pi\pi$ invariant mass distribution can be written as
\begin{align}
    \frac{\mathrm{d}\Gamma}{\mathrm{d}m_{\pi\pi}} = \frac{p_{\pi^-}p_{J/\psi}}{96 \pi^3 m_{\psi^\prime}^2} \sum_{\mathrm{spin}} \left| \mathcal{M} \right|^2,
    \label{Eq:pipidis}
\end{align}
where
\begin{align}
p_{J/\psi}=\frac{\sqrt{\lambda(m_{\psi^\prime}^2,E^2,m_{J/\psi}^2)}}{2m_{\psi^\prime}},\quad p_{\pi^-}=\frac{\sqrt{\lambda(E^2,M_{\pi}^2,M_{\pi}^2)}}{2E},
\end{align}
are the 3-momenta of the $J/\psi$ in the $\psi^\prime$ rest frame and the final-state $\pi^-$ in the $\pi\pi$ rest frame, respectively, $\sum_{\text{spin}}$ corresponds to the sum over the polarizations of the $\psi^\prime$ and $J/\psi$, and the amplitude $\mathcal{M}$ is given by Eq.~\eqref{Eq_MnnLO} for $\psi^\prime \to J/\psi \pi^0\pi^0$ and Eq.~\eqref{Eq_MccLO} for $\psi^\prime \to J/\psi \pi^+\pi^-$.

\begin{figure}[tb]
\centering
       
    {\includegraphics[width=0.68\linewidth]{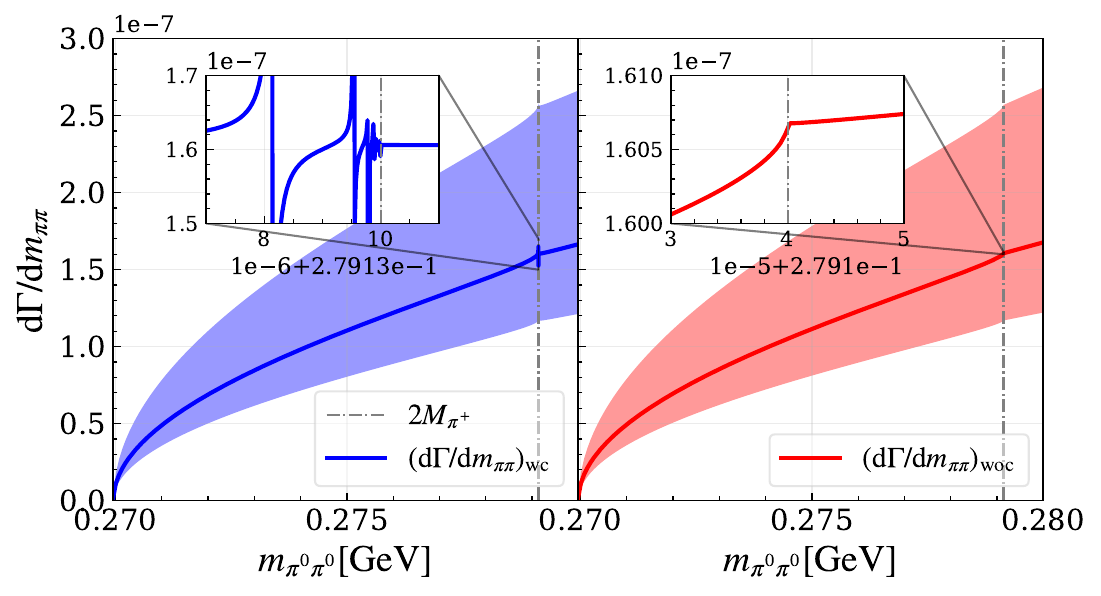}}
    
    \caption{The invariant mass distribution of the $\pi^0\pi^0$ pair for $\psi^\prime \to J/\psi \pi^0\pi^0$. The blue and red curves represent the distributions with and without the Coulomb effect, respectively, using the central values of $(a_0-a_2) M_{\pi^+}$ and $a_2 M_{\pi^+}$, $\alpha^{21}$ and $\kappa$. 
    The error bands are dominated by the errors of $\alpha^{21}$ and $\kappa$.
    The insets enlarge the line shape with and without Coulomb effects in the very near-threshold region. The gray vertical dot-dashed line denotes the $\pi^+\pi^-$ threshold. }
    \label{Figs_DWpi0pi0_CI}
\end{figure}
\begin{figure}[tb]
\centering
\includegraphics[width=0.5\linewidth]{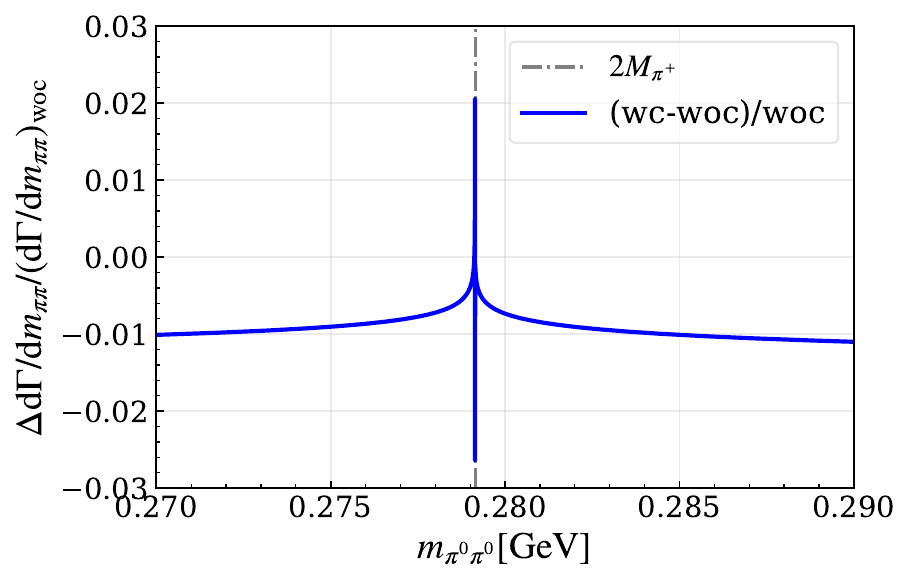}
\caption{Ratio of the difference $\Delta\mathrm{d}\Gamma/\mathrm{d} m_{\pi\pi}$ between the $\pi^0\pi^0$ invariant mass distributions with and without Coulomb interactions (denoted as ``wc'' and ``woc'', respectively) to the invariant mass distribution without Coulomb interactions. }
\label{Figs_Ratio}
\end{figure}

\section{Near-threshold line shapes}\label{Sec:Results}
With the parameters in the amplitude of $\psi^\prime \to J/\psi \pi\pi$ settled, we give predictions of the $\pi\pi$ invariant mass distributions.

In Fig.~\ref{Figs_DWpi0pi0_CI}, we compare the line shapes of the $\pi^0\pi^0$ distribution in the near-threshold region with ($(\mathrm{d}\Gamma/\mathrm{d} m_{\pi\pi})_{\mathrm{wc}}$, represented by the blue solid line) and without ($(\mathrm{d}\Gamma/\mathrm{d} m_{\pi\pi})_{\mathrm{woc}}$, obtained by setting $\kappa_c \to ik_c$~\cite{NA482:2005wht,Gevorkyan:2013tsa} in Eq.~\eqref{Eq:pipidis}, represented by the red solid line) Coulomb interactions. The cusp at the $\pi^+\pi^-$ threshold becomes more prominent when Coulomb interactions are included. 
The inset magnifies the line shapes in the immediate vicinity of the $\pi^+\pi^-$ threshold and reveals numerous peaks from the ground and excited states of pionium. Fig.~\ref{Figs_Ratio} shows the ratio of the difference $\Delta\mathrm{d}\Gamma/\mathrm{d} m_{\pi\pi}=(\mathrm{d}\Gamma/\mathrm{d} m_{\pi\pi})_{\mathrm{wc}}-(\mathrm{d}\Gamma/\mathrm{d} m_{\pi\pi})_{\mathrm{woc}}$ between the central values of the $\pi^0\pi^0$ spectrum with and without Coulomb interactions to the central value of the $\pi^0\pi^0$ spectrum $(\mathrm{d}\Gamma/\mathrm{d} m_{\pi\pi})_{\mathrm{woc}}$ without Coulomb interactions. The Coulomb corrections can change the magnitude of the cusp at the $\pi^+\pi^-$ threshold by approximately 2\%-3\%. Therefore, electromagnetic corrections should be included for precise extraction of the $\pi \pi$ scattering lengths in charmonium dipion transitions.

\begin{figure}[tb]
    \centering
    {\includegraphics[width=0.48\linewidth]{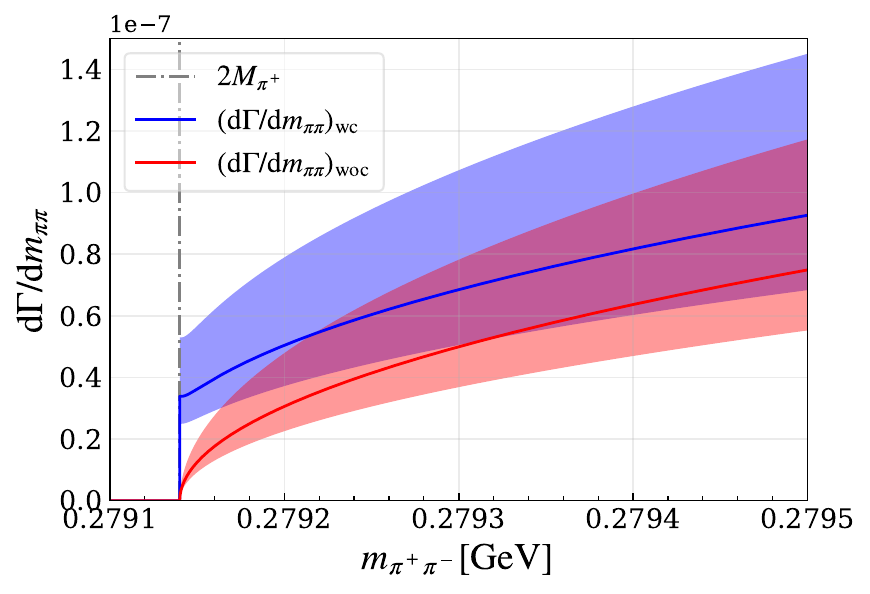}}
    
    \caption{The invariant mass distribution of the $\pi^+\pi^-$ pair for $\psi^\prime \to J/\psi \pi^+\pi^-$. The blue and red line shapes represent the distributions with and without the Coulomb interactions, respectively.} 

    \label{Figs_DWpicpic_CI}
\end{figure}

Fig.~\ref{Figs_DWpicpic_CI} presents a comparison of the line shapes for the $\pi^+\pi^-$ spectrum with (blue solid line) and without (red solid line) Coulomb interactions. The distribution with Coulomb interactions is nonzero at the threshold due to contributions from the pionium excited state poles in $W_C(E)$, as shown in the inset of Fig.~\ref{Figs_DWpi0pi0_CI}.

\section{Monte Carlo simulations}\label{Sec:MC}
\begin{figure*}[tb]
\centering
\includegraphics[width=\textwidth]{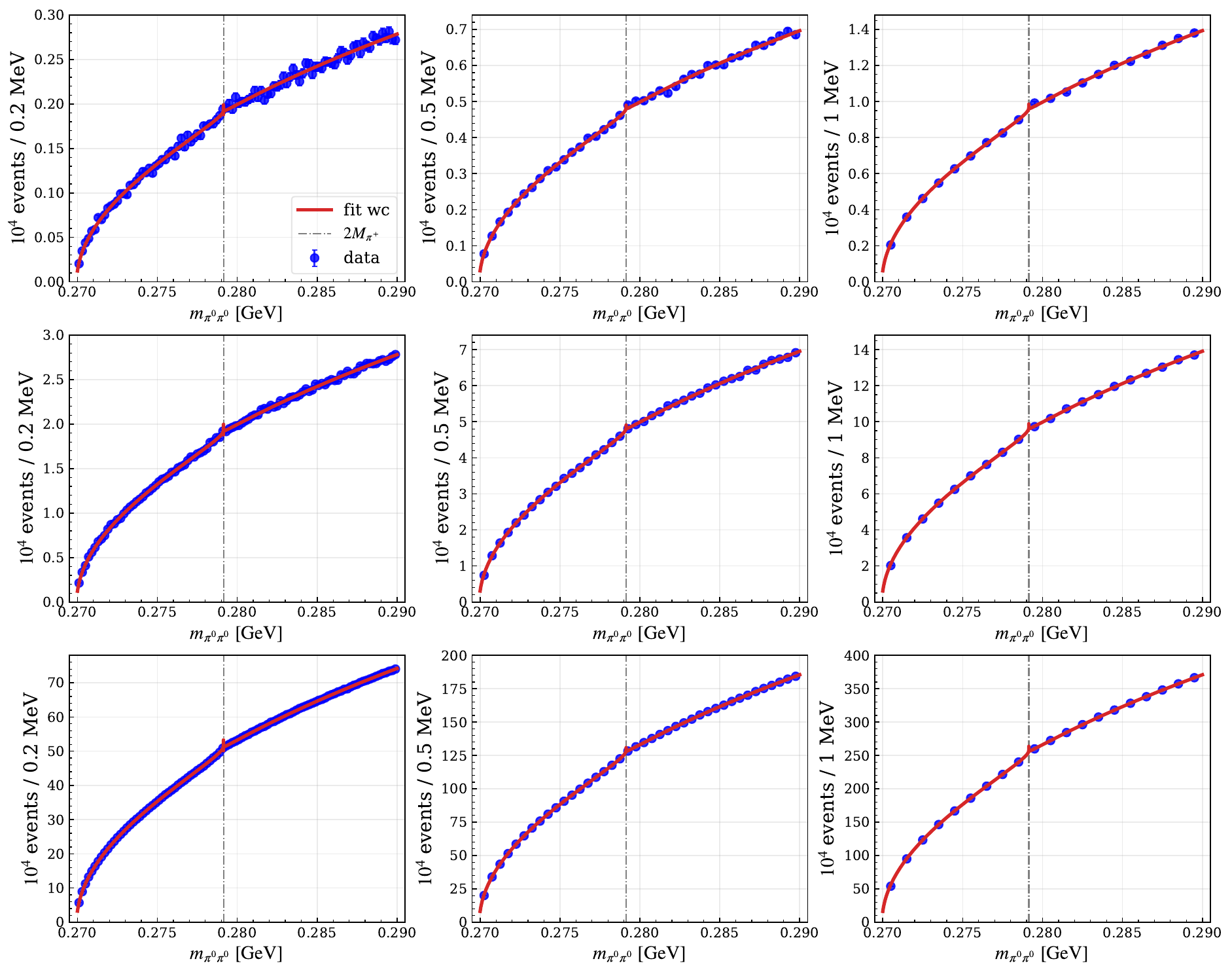}
\caption{Various sets of MC events that conform to the normalized $\pi^0\pi^0$ invariant mass distribution with Coulomb corrections and the best fits using the model with Coulomb interactions. The event numbers in the range $[0.270, 0.290]$ GeV for the first, second, and third rows are about $1.5 \times 10^5$, $1.5 \times 10^6$, and $4 \times 10^7$, respectively. The bin widths for the first, second, and third columns are 0.2, 0.5, and 1 MeV, respectively. The gray vertical dot-dashed line indicates the $\pi^+\pi^-$ threshold.}
\label{Figs_MCfit_wc}
\end{figure*}

In this section, we perform MC simulations to quantitatively estimate the precision with which electromagnetic corrections can influence the experimental extraction of $\pi\pi$ scattering lengths.
 
We use the von Neumann rejection method to generate random data points that conform to the normalized $\pi^0\pi^0$ distribution within the energy spectrum ranging from $0.270~\rm{GeV}$ to $0.290~\rm{GeV}$, as predicted by the NREFT in Eq.~\eqref{Eq:pipidis} with Coulomb interactions included. These data are generated using the central values $(a_0-a_2) M_{\pi^+}=0.2640$, $a_2 M_{\pi^+}=-0.0444$~\cite{Gasser:2009zz,Colangelo:2000jc,Colangelo:2001df}, $|\alpha^{21}|=1.18$, and $\kappa=0.26$~\cite{Wu:2024xwy}, and are subdivided into bins, with statistical uncertainties given by the square root of the event number in each bin. By varying the MC event numbers and bin widths, we aim to explore how the number of events and data binning \footnote{In principle, one also needs to consider the energy resolution by convolving the $\pi\pi$ invariant mass distribution with a Gaussian function.} influence the extent to which electromagnetic corrections affect the precision of scattering length extraction.

\begin{figure*}[tb]
\centering
\includegraphics[width=\textwidth]{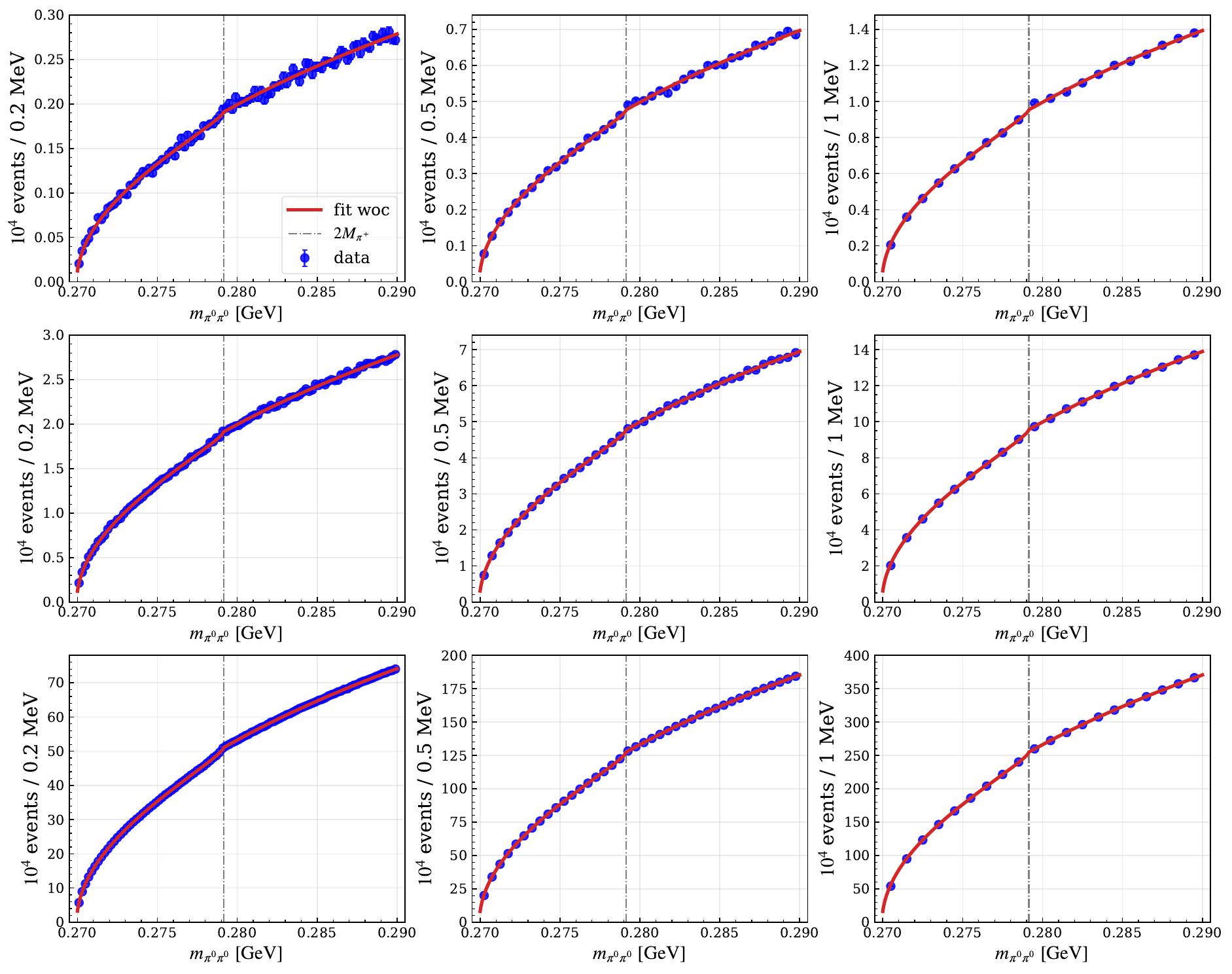}
\caption{Same data sets of MC events as in Fig.~\ref{Figs_MCfit_wc} and the best fits using the model without the Coulomb interactions. Other notations are same as those in Fig.~\ref{Figs_MCfit_wc}.}
\label{Figs_MCfit_woc}
\end{figure*}

Using the amplitude with $\pi\pi$ FSIs from Ref.~\cite{Wu:2024xwy}, one can estimate that the fraction of the $\pi^0\pi^0$ distribution below 0.29 GeV comprises about $3\times 10^{-4}$ of the total distribution over the entire phase space.
Considering the branching fraction of  $(18.2 \pm 0.5) \%$ for $\psi^\prime \to J/\psi \pi^0 \pi^0$~\cite{ParticleDataGroup:2024cfk}, the $2.7\times 10^9$ $\psi^\prime$ events collected at the BESIII experiment~\cite{BESIII:2024lks} corresponds to about $1.5 \times 10^5$ $\psi^\prime \to J/\psi \pi^0 \pi^0$ events for $m_{\pi^0\pi^0}\leq 0.290$~GeV.
We tested a number of different combinations of event numbers ($1.5 \times 10^5$, $1.5 \times 10^6$, and $4 \times 10^7$)\footnote{The last event number corresponds to the expected value at STCF~\cite{Achasov:2023gey}, which is more than the event number in the same region used in the NA48/2 analysis~\cite{Batley:2009ubw}.} and bin widths (0.1, 0.2, 0.5, 1, and 2~MeV) in the range of $[0.270,0.290]$ GeV, as shown in Figs.~\ref{Figs_MCfit_wc} and~\ref{Figs_MCfit_woc}.

The $\pi^0\pi^0$ invariant mass distribution was then fitted to the same MC data sets using the equation
\begin{align}
    \mathrm{events} = \mathcal{N} \times \frac{\mathrm{d}\Gamma [\psi^\prime \to J/\psi \pi^0\pi^0]}{\mathrm{d}m_{\pi^0\pi^0}}, 
\end{align}
with and without Coulomb interactions, where the free parameters include an overall normalization factor $\mathcal{N}$ and scattering lengths $(a_0-a_2) M_{\pi^+}$ and $a_2 M_{\pi^+}$. 

The best fits using the model with Coulomb interactions are represented by the red lines in Fig.~\ref{Figs_MCfit_wc}, and the red lines in Fig.~\ref{Figs_MCfit_woc} show the best fitted results obtained using the model without Coulomb interactions. The values of the scattering lengths $(a_0-a_2) M_{\pi^+}$ extracted from the fits are collected in Table~\ref{Tab_a0ma2}. The values outside and inside parentheses in the table correspond to the presence or absence of Coulomb interactions in the fitting model, respectively. These fitted results are not guaranteed to be identical to the input ($(a_0-a_2)M_{\pi^+}=0.2640$) due to random fluctuations during the data generation process. One can see that the precision of the extracted value can reach $(4-6)\%$ for $1.5 \times 10^5$, $(2-3)\%$ for $1.5 \times 10^6$, and $(0.3-0.6)\%$ for $4 \times 10^7$ events in the range of $[0.270,0.290]$ GeV. For larger event numbers, the precision improves by a factor of approximately $\sqrt{N/N^\prime}$, where $N^\prime$ and $N$ represent the new and old event numbers, respectively, and the central values of the extracted scattering lengths are closer to the input value. For different bin widths, the extraction precision remains remarkably consistent, as found in Ref.~\cite{Liu:2012dv}.

The scattering lengths extracted using the two amplitudes with and without Coulomb interactions are similar, 
however, the central values exhibit a difference of 1\%-5\%. 
When the extraction precision approaches 1\% (number of events $\sim \mathcal{O}(10^7)$ in our simulation), 
the input value of the scattering length $(a_0-a_2) M_{\pi^+}$ could fall outside the error region of the result extracted using the amplitude without Coulomb interactions, as can be seen from Table~\ref{Tab_a0ma2}. 

\begin{table}[tb]
\centering
\caption{\label{Tab_a0ma2} Results of fitting to various sets of MC data. The extracted scattering lengths are given in units of $M_{\pi^+}^{-1}$. The values outside and inside the parentheses correspond to the presence or absence of Coulomb interactions in the fitting models. The uncertainties only reflect the statistical errors in the fit.}
\begin{ruledtabular}
\begin{tabular}{cccccc}

        Bin width
        & Events
        & $1.5 \times 10^5$
        & $1.5 \times 10^6$
        & $4 \times 10^7$
        \\[3pt]     
\hline
  \multirow{2}{*}{0.1 MeV} 
  &$\chi^2/$d.o.f &0.90(0.91) &1.10(1.10) &0.97(1.14) \\
  \cline{2-5}
  &$a_0-a_2$ 
  &$0.270_{-0.012}^{+0.011} (0.260 \pm 0.014)$ 
  &$0.265_{-0.005}^{+0.004} (0.268 \pm 0.005)$ 
  &$0.2646_{-0.0010}^{+0.0009} (0.2674 \pm 0.0016)$ \\
  \hline
  \multirow{2}{*}{0.2 MeV} 
  &$\chi^2/$d.o.f &0.90(0.91) &1.13(1.15) &0.85(1.19) \\
   \cline{2-5}
  &$a_0-a_2$ 
  &$0.268_{-0.010}^{+0.011} (0.259_{-0.012}^{+0.013})$ 
  &$0.265 \pm 0.005 (0.268 \pm 0.008)$ 
  &$0.2645 \pm 0.0009 (0.2674_{-0.0016}^{+0.0015})$ \\
  \hline
  \multirow{2}{*}{0.5 MeV} 
  &$\chi^2/$d.o.f &0.60(0.63) &0.93(0.93) &0.96(1.65) \\
   \cline{2-5}
  &$a_0-a_2$ 
  &$0.262_{-0.012}^{+0.011} (0.256 \pm 0.014)$ 
  &$0.265_{-0.004}^{+0.005} (0.268_{-0.007}^{+0.008})$ 
  &$0.2646_{-0.0009}^{+0.0008} (0.2673 \pm 0.0016)$ \\
  \hline
  \multirow{2}{*}{1 MeV} 
  &$\chi^2/$d.o.f &0.54(0.59) &0.57(0.58) &0.91(2.33) \\
   \cline{2-5}
  &$a_0-a_2$ 
  &$0.264 \pm 0.010 (0.255 \pm 0.014)$ 
  &$0.266 \pm 0.005 (0.269 \pm 0.008)$ 
  &$0.2647_{-0.0011}^{+0.0008} (0.2674 \pm 0.0016)$ \\
  \hline
  \multirow{2}{*}{2 MeV} 
  &$\chi^2/$d.o.f &0.50(0.63) &0.29(0.30) &1.48(3.31) \\
   \cline{2-5}
  &$a_0-a_2$ 
  &$0.266 \pm 0.010 (0.252_{-0.015}^{+0.014})$ 
  &$0.266 \pm 0.005 (0.269_{-0.008}^{+0.007})$ 
  &$0.2648_{-0.0009}^{+0.0010} (0.2671 \pm 0.0016)$ \\

\end{tabular}
\end{ruledtabular}
\end{table}

\section{Summary}\label{Sec:Summary}

In this work, we studied the near-threshold $\pi\pi$ invariant mass distributions in $\psi^\prime \to J/\psi \pi\pi$ within the NREFT framework, incorporating effects from Coulomb interactions. 
Our results show that the cusp structure in the $\pi^0\pi^0$ invariant mass distribution at the $\pi^+\pi^-$ threshold becomes slightly more prominent when Coulomb interactions are included. The Coulomb corrections can change the magnitude of the cusp by about (2-3)\%, indicating the necessity of including the electromagnetic corrections for accurately extracting the $\pi\pi$ scattering lengths from the threshold cusp. 
To evaluate the impact of electromagnetic interactions on the extraction of scattering lengths, we generated artificial data using MC simulation and performed fits with $(a_0-a_2) M_{\pi^+}$ and $a_2 M_{\pi^+}$ as free parameters.
We found that neglecting the Coulomb corrections can lead to an overestimation of the scattering lengths, especially for experiments with high energy resolution. This level of precision corresponds to about 20 million $\psi'$ events in the considered region $m_{\pi\pi}\in [0.270,0,290]$~GeV. For data sets with significantly fewer events, omitting electromagnetic corrections in the extraction of scattering lengths is likely acceptable. 
Our approach provides a reliable framework for the precise description of the near-threshold $\pi^0\pi^0$ invariant mass distribution in $\psi^\prime \to J/\psi \pi^0\pi^0$, applicable to data from BESIII and future STCF experiments.
Deviations from the predicted distribution may indicate the necessity of including additional effects, such as the exchange of isospin-1 $Z_c$ structures~\cite{Guo:2004dt} and the $J/\psi \pi$ FSI.

\begin{acknowledgments}
We would like to thank Feng-Kun Guo valuable suggestions, careful reading, and modifications of this manuscript, Shuang-Shi Fang, Mao-Jun Yan, and Yin Cheng for helpful discussions, and thank Bing Wu and Xiang-Kun Dong for providing data about the $\psi^\prime \to J/\psi\pi\pi$ vertex. This work is supported in part by the National Key R\&D Program of China under Grant No. 2023YFA1606703; by the National Natural Science Foundation of China (NSFC) under Grants No.~12125507, No.~12447101, No.~12475081, and No.~12361141819;  by the Natural Science Foundation of Shandong province under Grants No.~ZR2025MS04, and No.~ZR2022ZD26; by Taishan Scholar Project of Shandong Province under Grant No.~tsqn202103062; by the Chinese Academy of Sciences under Grant No.~YSBR-101; and by the Postdoctoral Fellowship Program of China Postdoctoral Science Foundation under Grant No. 2025M773427.
\end{acknowledgments}

\bibliography{refs}
\end{document}